\documentclass[lettersize,journal]{IEEEtran}
\usepackage{cite}

\ifCLASSINFOpdf
  \usepackage[pdftex]{graphicx}
\else
\fi
\usepackage{amsmath}
\usepackage{amsfonts}
\usepackage{amssymb}
\usepackage{color}
\usepackage{algorithmic}
\usepackage{algorithm}

\usepackage{array}

\ifCLASSOPTIONcompsoc
  \usepackage[caption=false,font=normalsize,labelfont=sf,textfont=sf]{subfig}
\else
  \usepackage[caption=false,font=footnotesize]{subfig}
\fi
\usepackage{fixltx2e}

\usepackage{stfloats}
\usepackage{url}

\newcommand{\norm}[1]{\left\lVert#1\right\rVert}	
\usepackage{amsmath, amsthm, amssymb}

\usepackage{booktabs}
\usepackage{varwidth}
\usepackage{comment}
\usepackage{xpatch}
\usepackage{lipsum}  
\makeatletter
\xpatchcmd{\proof}{\@addpunct{.}}{\@addpunct{:}}{}{}

\makeatother
\begin{document}

%
\title{Electromagnetic Interference Cancellation for RIS-Assisted Communications}
%
%
%
\author{Aymen~Khaleel, and
        Ertugrul~Basar,~\IEEEmembership{Fellow,~IEEE}
\thanks{This work was supported by the Scientific and Technological Research Council of Turkey (TUBITAK) under Grant 120E401.
	
The authors are with the Communications Research and Innovation Laboratory (CoreLab), Department of Electrical and Electronics Engineering, Ko\c{c} University, Sariyer 34450, Istanbul, Turkey. (e-mail: akhaleel@ku.edu.tr, ebasar@ku.edu.tr).}
}
\maketitle

\begin{abstract}
 Reconfigurable intelligent surface (RIS)-empowered communication is an emerging technology that has recently received growing attention as a potential candidate for next-generation wireless communications. Although RISs have shown the potential of manipulating the wireless channel through passive beamforming, it is shown that they can also bring undesired side effects, such as reflecting the electromagnetic interference (EMI) from the surrounding environment to the receiver side. In this study, we propose a novel EMI cancellation scheme to mitigate the impact of the EMI by exploiting its special time-domain structure and considering a clever passive beamforming method at the RIS. Compared to its benchmark, computer simulations show that the proposed scheme achieves superior performance in terms of the average signal-to-interference-plus-noise ratio (SINR) and outage probability (OP), especially when the EMI power is comparable to the power of the information signal impinging on the RIS surface.
\end{abstract}

%
\IEEEpeerreviewmaketitle
%
%
%
%
\section{Introduction} 
\IEEEPARstart{R}{econfigurable} intelligent surfaces (RISs) have recently received growing attention due to their unique capabilities and wide applications in wireless communications systems \cite{Transmission_conference}. In particular, RISs have been integrated to the vast majority of existing wireless communications systems such as multiple-input multiple-output (MIMO) \cite{AK1}, non-orthogonal multiple access (NOMA) \cite{AK2}, \cite{mahmoud}, and many others. An RIS is mainly used to passively steer the transmitted signal by aligning a large number of its reflections (from the RIS surface) such that they constructively combine at the receiver side \cite{Transmission_conference}. However, due to its inherent physical nature, an RIS cannot selectively reflect the signals impinging its surface; therefore, it also reflects undesired signals from the surrounding environment, causing electromagnetic interference (EMI) at the receiver side \cite{EMI_Emil}. So far, few works have shed light on this problem and took it into consideration while designing RIS-assisted systems. In particular, the authors in \cite{EMI_Emil} provided a mathematical model for an RIS-assisted single-input single-output (SISO) system under EMI interference, where they modeled the EMI as a complex Gaussian random vector with zero mean and a variance corresponding to the EMI power. The authors of \cite{EMI_Emil} also showed that the EMI could change the signal-to-noise ratio (SNR) scaling behavior where, due to the EMI, the SNR grows linearly with the RIS size and not quadratically as in the conventional case without EMI \cite{ris-snr-scale}-\cite{snr-sqr-law}. This proves that the EMI has a critical negative impact on the performance of RIS-assisted systems by significantly limiting the RIS passive beamforming gain. The EMI's main possible sources are other transmitters in the RIS surroundings where their signals are unintentionally reflected from the RIS to the desired user. In \cite{emi-relay}, the authors compared RIS-assisted and decode-and-forward (DF) relaying systems under EMI in terms of minimizing the total transmit power. In \cite{emi-urllc}, the performance of an RIS-assisted ultra-reliable low-latency communication (URLLC) system is investigated for multi-user scenario in the presence of EMI. In \cite{emi-phys}, the authors investigated the EMI effect on the achievable secrecy performance in RIS-assisted communications systems. In \cite{emi-multipair}, the performance of a multi-pair full-duplex (FD) two-way communication system is considered under RIS hardware impairments, spatial correlation, and EMI. In \cite{RIS-Interf}, the authors used the RIS to null the mutual interference between single-antenna transceivers in a multiuser system by properly designing its phase shifts. Note that this kind of interference differs from the one considered in \cite{EMI_Emil}, as the latter is uncontrollable, and the associated channel cannot be estimated. However, to the best of the authors' knowledge, no work in the RIS literature has yet investigated the problem of mitigating the EMI in RIS-assisted communications.

Against this background, we propose a novel EMI cancellation scheme to mitigate the EMI impact and enhance the signal-to-interference-plus-noise ratio (SINR) at the receiver side. In particular, by exploiting the special structure of the RIS channels and the time-domain behavior of the EMI, we trade off the passive beamforming gain in favor of canceling the EMI through the proper design of the RIS phase shifts. Computer simulations with different system settings show that the proposed scheme has a superior performance in terms of the average SINR and outage probability. 

The rest of the paper is organized as follows. In Section \ref{sec:emi-sysmdl}, we introduce the system model and explain the proposed EMI cancellation scheme. In Section \ref{sec:emi-op}, we give outage probability analysis. Section \ref{sec:emi-sim} provides computer simulations followed by conclusions in Section \ref{sec:emi-conc}.
\section{System Model}\label{sec:emi-sysmdl}
\begin{figure}[t]
	\centering
	\includegraphics[width=80mm]{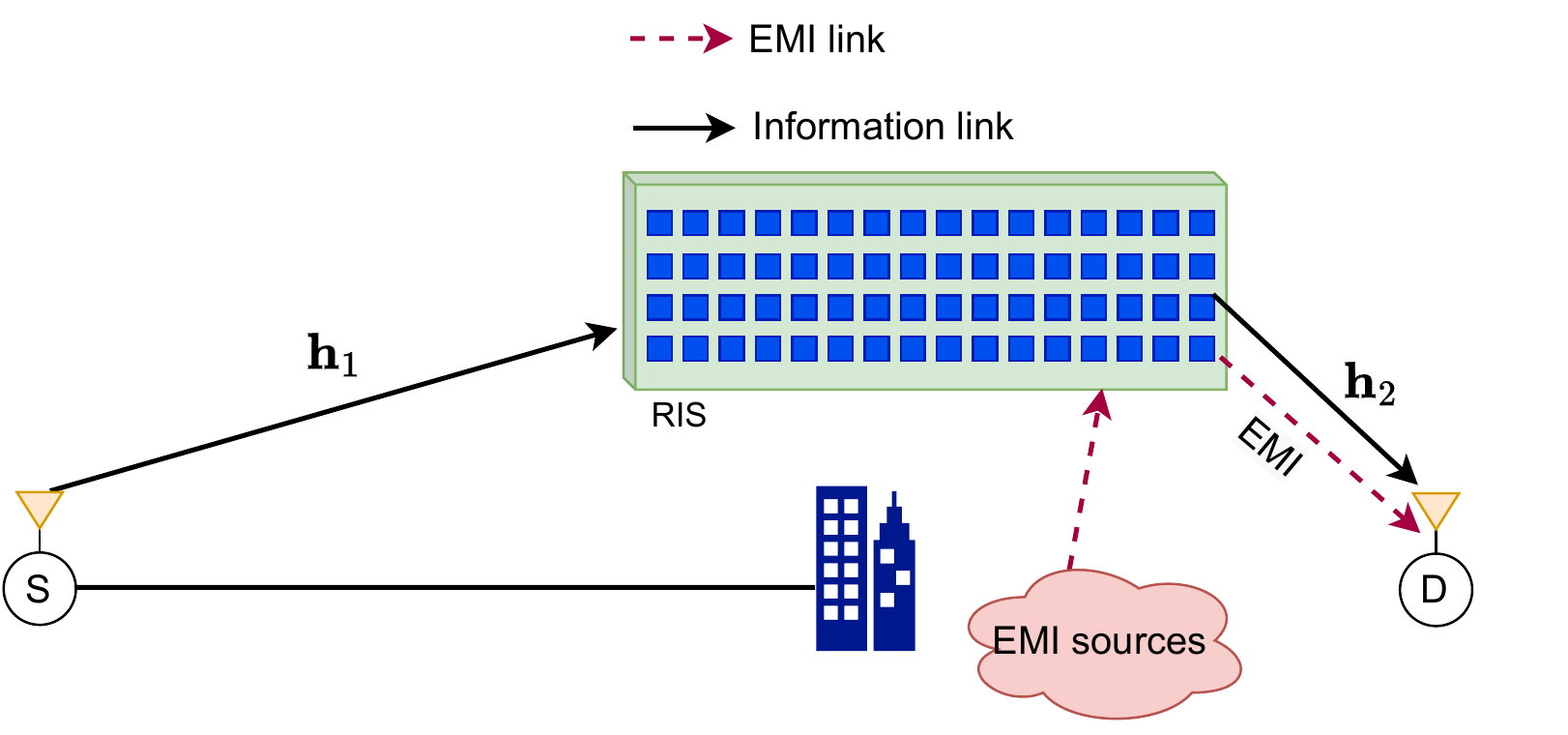}
\caption{RIS-assisted SISO communications system under EMI.}\label{fig:system_block}
\vspace{-0.5cm}
\end{figure}
In this section, we first introduce the system model then we explain the proposed EMI cancellation scheme in detail. We consider an RIS-assisted SISO system, as shown in \mbox{Fig. \ref{fig:system_block}}, where an RIS consisting of $N$ elements is employed. The direct communication link between the source (S) and the destination (D) is assumed to be blocked due to obstacles, and the reflection link over the RIS is the only available link \footnote{This makes the impact of the EMI more evident. Otherwise when there is a powerful direct link, the use of RIS becomes insufficient, unless it is extremely large to make a noticeable difference for the SNR.}. Furthermore, in addition to the desired transmitted signal, the RIS is assumed to reflect EMI signals from local interference sources close to D \cite{EMI_Emil}. Accordingly, considering only the first reflection from the RIS, at any given time slot $t$, the received signal at D can be given as \cite{EMI_Emil}
\begin{align}
y_t=\mathbf{h}_{2,t}^H\mathbf{\Theta}_t^H\mathbf{h}_{1,t}\sqrt{P}s_t+\mathbf{h}_{2,t}^H\mathbf{\Theta}_t^H\mathbf{n}_t+w_t,\label{eq:rx}
\end{align}   
where $s_t$ is the transmitted message at time slot $t$ with a transmit power $P$. Here,  $\mathbf{\Theta}_t\in\mathbb{C}^{N\times N}$ is a diagonal matrix containing the RIS reflection coefficients with $\lfloor\mathbf{\Theta}_t\rfloor_i=\eta_{i,t}e^{j\theta_{i,t}}$, where $\lfloor\mathbf{z}\rfloor_i$ denotes the $i$-th entry of the vector $\mathbf{z}$, $\eta_{i,t}=1$ and $\theta_{i,t}\in[0,2\pi),\forall i,t$, are the $i^{th}$ element applied reflection amplitude and phase shift at time slot $t$, respectively. $\mathbf{h}_{1,t}$ and $\mathbf{h}_{2,t}\in\mathbb{C}^{N\times 1}$ are the S-RIS and RIS-D channel vectors at time slot $t$, respectively. $\mathbf{h}_{k,t}\sim\mathcal{CN}(\mathbf{0},A\beta_k\mathbf{R}),k\in\{1,2\}$ \cite{spatial-corr}, where $\beta_k$ is the path gain, $A=d_Hd_V$ is the RIS element area with $d_V$ and $d_H$ are the length and width, respectively, and $\mathcal{CN}$ represents complex Gaussian random distribution with $\mathbf{R}$ denoting the RIS correlation matrix and $\mathbf{0}$ denoting $N$-dimensional
all-zeros column vector. Assuming isotropic conditions for the EMI, $\mathbf{n}\in\mathbb{C}^{ N\times 1}$ is the EMI vector, where $\mathbf{n}\sim\mathcal{CN}(\mathbf{0},A\sigma^2\mathbf{R})$ with $\sigma^2$ is the EMI power \cite{EMI_Emil}. Accordingly, the ratio of the signal power to the EMI power at each RIS element can be expressed as $\rho=\beta_1P/\sigma^2$. $w\sim\mathcal{CN}(0,\sigma_w^2)$ is the additive white Gaussian noise (AWGN) sample with zero mean and variance $\sigma_w^2$.

\begin{figure}[t]
	\centering
	\includegraphics[width=46mm]{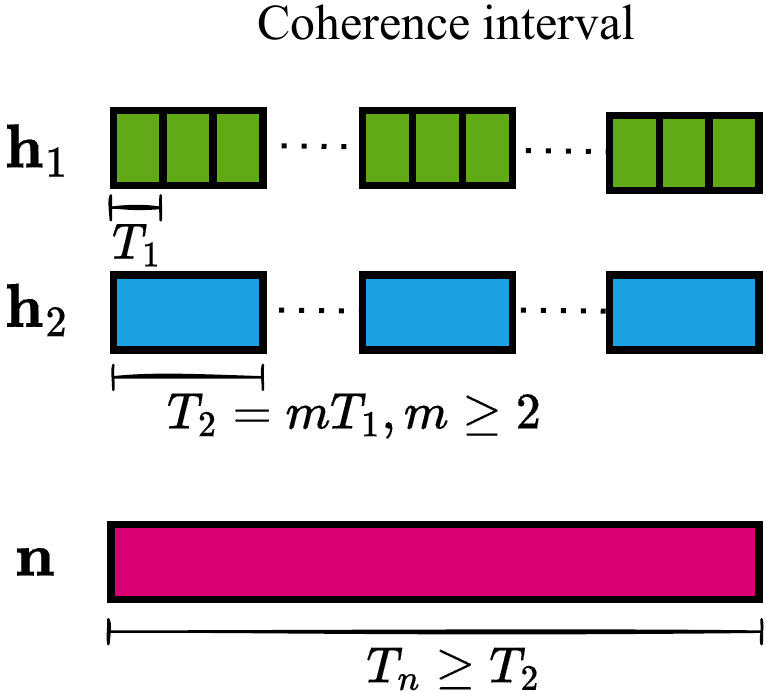}
	\caption{The coherence intervals (CIs) $T_1$ of $\mathbf{h}_1$, $T_2$ of $\mathbf{h}_2$, and $T_n$ of $\mathbf{n}$ are illustrated relative to each other, where the random realization of each channel/EMI remains constant over its corresponding CI.}\label{fig:CIs}
	\vspace{-0.5cm}
\end{figure}

To mitigate the high path loss associated with the reflection link, the RIS needs to be deployed close to S or D \cite{ris-position}, where its position can be optimized accordingly \cite{ris-position2}. Consequently and without loss of generality, we assume the RIS to be deployed close to D, which typically can be at a fixed location like a base station, while S can be a mobile user equipment. In this way, the RIS-D channel $\mathbf{h}_2$ changes less frequently compared to the other channels, where the statistical channel state information (S-CSI) dominates the instantaneous CSI (I-CSI) in terms of the phase shift design \cite{H2_slow1}, \cite{H2_slow2}. On the other side, depending on the nature of the interference source and the considered system setup, the realizations of the EMI vector $\mathbf{n}$ can change faster or slower than those of the other channels. While the realizations of EMI cannot be generally assumed to be changing slower than those of other channels associated with the considered system, we consider here a specific, yet, realistic case where the coherence interval (CI) for $\mathbf{n}$ spans several CIs of the other channels. This case can be observed when the EMI source is a secondary transmitter in the network, like a backscatter or Internet of Things (IoT) device with a low transmission rate \cite{EMI_slow}. Fig. \ref{fig:CIs} illustrates the considered CIs for different channels with respect to the EMI, where the realizations of $\mathbf{n}$ and $\mathbf{h}_2$  change slower than that of $\mathbf{h}_1$, due to the reasons explained earlier.
We consider a flat-fading channel where the channel realization remains constant within each CI and independent from the realizations at other CIs. In particular, as shown in Fig. \ref{fig:CIs}, the realizations of $\mathbf{h}_1$ and $\mathbf{h}_2$ remain constant over the time durations of $T_1$ and $T_2$, respectively, while the realization of $\mathbf{n}$ remains constant for a duration greater than $T_2$. In order to satisfy the slow fading condition, $s$ is transmitted at each time slot of duration $T_1$, where, in addition to the EMI realization, all channels remain constant.

From \eqref{eq:rx}, the SINR at time slot $t$ ($\text{SINR}_t$) is given by\footnote{Note that, as stated in \cite{EMI_Emil}, the EMI interference is treated as a sort of additive noise. However, here, we refer to it as interference for notational convenience.}
\begin{align}
	\text{SINR}_t=\frac{P|\mathbf{h}_{2,t}^H\mathbf{\Theta}_t^H\mathbf{h}_{1,t}|^2}{A\sigma^2\mathbf{h}_{2,t}^H\mathbf{\Theta}_t^H\mathbf{R}\mathbf{\Theta}_t\mathbf{h}_{2,t}+\sigma_w^2},\label{eq:snr1}
\end{align}
where $\text{E}[|\mathbf{h}_{2,t}^H\mathbf{\Theta}_t^H\mathbf{n}_t|^2]=A\sigma^2\mathbf{h}_{2,t}^H\mathbf{\Theta}_t^H\mathbf{R}\mathbf{\Theta}_t\mathbf{h}_{2,t}$ \cite{EMI_Emil}, and $\text{E}[\cdot]$ is the expectation operator. It can be noted from \eqref{eq:snr1} that the presence of EMI has a negative impact on the SINR. In particular, due to the EMI, it has been shown in \cite{EMI_Emil} that the SINR grows linearly with $N$ and not quadratically as in the conventional case without EMI \cite{ris-snr-scale} \cite{snr-sqr-law}. 
\subsection{Proposed EMI Cancellation Scheme}
To mitigate the EMI impact on the D side, we exploit the system settings explained earlier to propose a novel interference cancellation scheme. Specifically, since the EMI realization is correlated only with $\mathbf{\Theta}_t$ and $\mathbf{h}_2$, and not with $\mathbf{h}_1$, the proposed scheme works within the CI of $\mathbf{h}_2$ ($T_2$), which can be explained as follows.

Consider the transmission of $m$ symbols within a single CI of $\mathbf{h}_2$. Thus, $T_2$ is divided into $m$ time slots, each with a duration of $T_1$ ($T_2=mT_1,m \geq 2$), where a single symbol $s$ is transmitted at each $T_1$. Accordingly, the received signal within first time slot of $T_2$ can be obtained by rewriting \eqref{eq:rx} as
\begin{align}
	y_1=&\left(\sum_{i=1}^{N}\left|\lfloor\mathbf{h}_{1,1}\rfloor_i\right|\left|\lfloor\mathbf{h}_{2,1}\rfloor_i\right|\right)\sqrt{P}s_1+\mathbf{h}_{2,1}^H\mathbf{\Theta}_1^H\mathbf{n}_1\nonumber\\ 
	&+w_1,\label{eq:rx2}
\end{align}
where, in order to maximize the SNR, the RIS phase shifts are adjusted such that $\theta_{i,1}=\arg(\lfloor\mathbf{h}_{1,1}\rfloor_i(\lfloor\mathbf{h}_{2,1}\rfloor_i)^H),\forall i$ \cite{EMI_Emil}. By letting $B=\sum_{i=1}^{N}\left|\lfloor\mathbf{h}_{1,1}\rfloor_i\right|\left|\lfloor\mathbf{h}_{2,1}\rfloor_i\right|$, the transmitted symbol can be detected at D using a maximum likelihood detector, as follows
\begin{align}
\hat{s}_1=\underset{s\in\mathcal{S}}{\arg\min}\left|y_1-Bs\right|^2,
\end{align}
where $\mathcal{S}$ is the set of all possible transmitted symbols. Accordingly, from \eqref{eq:rx2}, the 
SINR to detect $s_1$ can be obtained as \cite{EMI_Emil}
\begin{align}
	\text{SINR}_1=\frac{PB^2}{A\sigma^2\mathbf{h}_{2,1}^H\mathbf{\Theta}_1^H\mathbf{R}\mathbf{\Theta}_1\mathbf{h}_{2,1}+\sigma_w^2}.\label{eq:snr-emi}
\end{align}
Next, we extract the EMI sample from $y_1$ as follows
\begin{align}
\tilde{E}&=y_1-B\hat{s}_1\\ \nonumber
&=\mathbf{h}_{2,1}^H\mathbf{\Theta}_1^H\mathbf{n}_1+I+w_1,
\end{align}
where $I=B(s_1-\hat{s}_1)$ corresponds to the residual interference due to unsuccessful detection of $s_1$.

At the remaining time slots within $T_2$, $t^{'}\in\{2, 3, ..., m\}$, the received signal can be obtained by rewriting \eqref{eq:rx} as follows
 \begin{align}
 	y_{t^{'}}=\mathbf{h}_{2,1}^H\mathbf{\Theta}_{t^{'}}^H\mathbf{h}_{1,t^{'}}\sqrt{P}s_{t^{'}}+\mathbf{h}_{2,1}^H\mathbf{\Theta}_{t^{'}}^H\mathbf{n}_1+w_{t^{'}},\label{eq:rx_t2}
 \end{align} 
where, as illustrated in Fig. \ref{fig:CIs}, $\mathbf{h}_{2,t^{'}}=\mathbf{h}_{2,1}$ and $\mathbf{n}_{t^{'}}=\mathbf{n}_1$, $\forall t^{'}\in\{2, 3, ..., m\}$. In order to cancel the EMI signals at D in the remaining time slots $t^{'}$ within $T_2$, we destructively add the EMI sample $\tilde{E}$ (obtained at the first time slot) to $y_{t^{'}}$, as follows
	\begin{align}
	\tilde{y}_{t^{'}}&=y_{t^{'}}+\tilde{E}\nonumber\\
	&=\mathbf{h}_{2,1}^H\mathbf{\Theta}_{t^{'}}^H\mathbf{h}_{1,t^{'}}\sqrt{P}s_{t^{'}}+\mathbf{h}_{2,1}^H(\mathbf{\Theta}_{t^{'}}^H+\mathbf{\Theta}_1^H)\mathbf{n}_1\nonumber\\
	&+I+v_{t^{'}},
\label{eq:rx_E}
\end{align}
where $v_{t^{'}}=w_1+w_{t^{'}}$. In order to fully eliminate the EMI sample, the RIS phase shifts can be adjusted such that $\mathbf{\Theta}_{t^{'}}=e^{j\pi}\mathbf{\Theta}_1, \forall t^{'}$, which comes at the expense of having no beamforming gain. Accordingly, the SINR at time slot $t^{'}\in\{2, 3, ..., m\}$ can be obtained as 
\begin{align}
	\text{SINR}_{t^{'}}=\frac{P|\mathbf{h}_{2,1}^H\mathbf{\Theta}_{t^{'}}^H\mathbf{h}_{1,t^{'}}|^2}{|I|^2+\sigma_{v}^2},\label{eq:snr-no-emi}
\end{align}
where $\sigma_{v}^2=2\sigma_w^2$ is the variance of $v_{t^{'}}$, which corresponds to the sum of the variances of the two independent random variables (RVs) $w_{1}$ and $w_{t^{'}}$.

By considering \eqref{eq:snr-emi} and \eqref{eq:snr-no-emi} together, it can be noted that, in terms of the RIS phase shift design, there is a trade-off between achieving passive beamforming gain and fully eliminating the EMI at D. In our computer simulations, we show that at high EMI levels, sacrificing the beamforming gain in favor of eliminating the EMI provides better results in terms of the average SINR and outage probability performance.
\section{Outage Probability Analysis}\label{sec:emi-op}
In this section, we provide the outage probability analysis to characterize the performance of the proposed EMI cancellation scheme, as follows.

\begin{figure}[t]
	\centering
	\includegraphics[width=60mm]{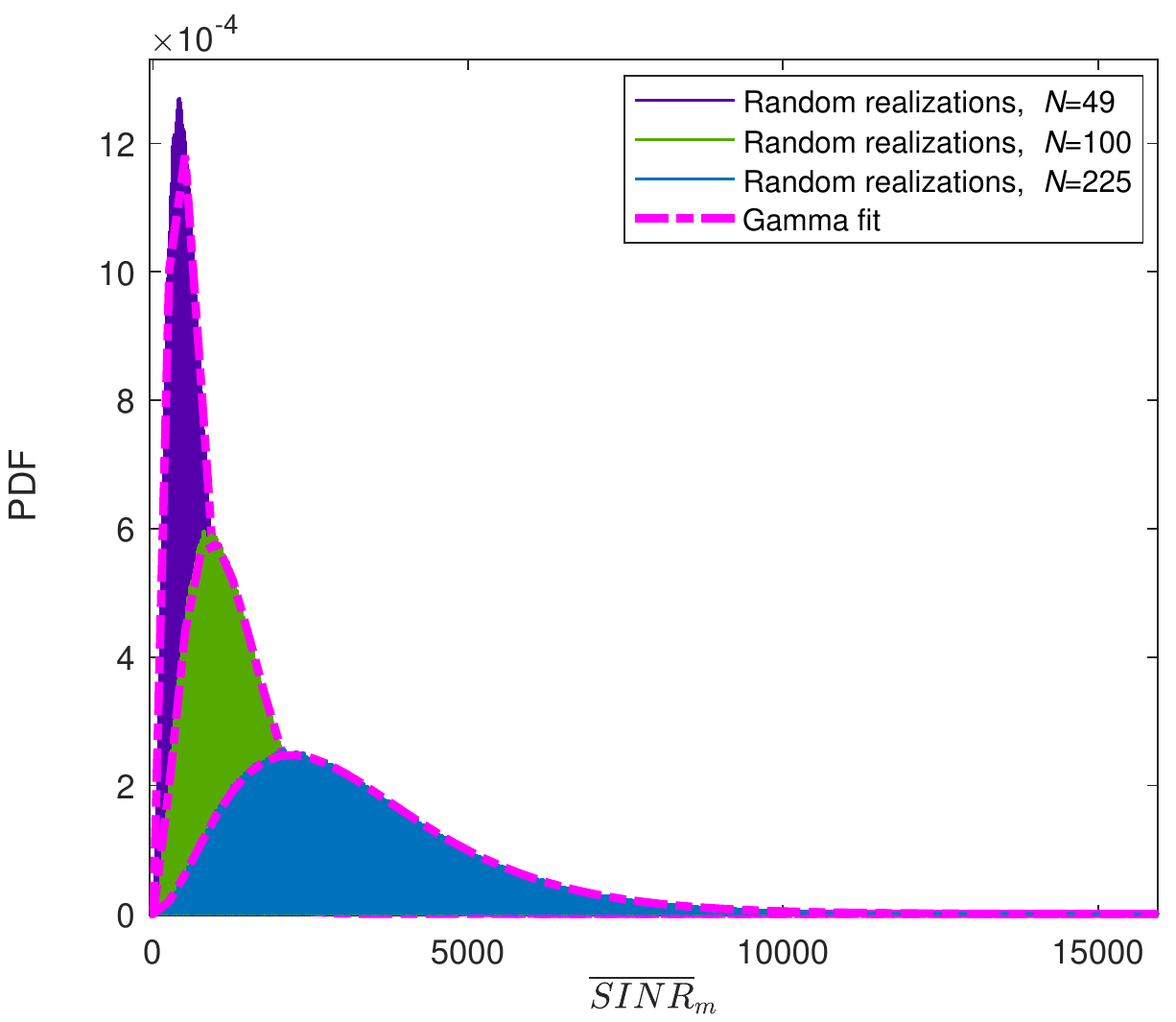}
	\caption{Fitting the distribution of $\overline{\text{SINR}}_m$ to Gamma distribution, for $\rho=0$ dB.}
	\label{fig:sinr-fit}
	\vspace{-0.5cm}
\end{figure}

Let $\text{SINR}_m$ denotes the average SINR over $m$ CIs such that
\begin{align}
\text{SINR}_m=\frac{1}{m}\sum_{t=1}^{m}\text{SINR}_t,\label{eq:sinr-ave}
\end{align}
where, for $t=1$ we have $\text{SINR}_1$ given in \eqref{eq:snr-emi}, and for $t>1$ we have $t=t^{'}\in\{2, 3, ..., m\}$ and $\text{SINR}_{t^{'}}$ is given in \eqref{eq:snr-no-emi}. Accordingly, for a given $\text{SINR}_m$ threshold $r$, the outage probability  (OP) can be obtained as
\begin{align}
P_{out}&=P\left(\text{SINR}_m<r\right)\nonumber\\
&=P\left(\overline{\text{SINR}}_m<\frac{r}{P}\right)\nonumber\\
&=F_{\overline{\text{SINR}}_m}\left(\frac{r}{P}\right),\label{eq:op}
\end{align}
where $\overline{\text{SINR}}_m=\text{SINR}_m/P$ and $F_{\overline{\text{SINR}}_m}(x)$ is the cumulative distribution function (CDF) of $\overline{\text{SINR}}_m$.
It can be noted from \eqref{eq:snr-emi} and \eqref{eq:snr-no-emi} that $\overline{\text{SINR}}_m$ involves the product and division of multiple correlated RVs, resulting in a very challenging task to obtain its exact probability density function (PDF). Therefore, we use a semi-analytical approach to obtain the PDF of $\overline{\text{SINR}}_m$ using the Distribution Fitting Tool in MATLAB. As shown in Fig. \ref{fig:sinr-fit}, the PDF of $\overline{\text{SINR}}_m$ perfectly matches the one of Gamma distribution. Consequently, $P_{out}$ can be obtained as the CDF of $\overline{\text{SINR}}_m$, which is given by
\begin{align}
F_{\overline{\text{SINR}}_m}\left(\frac{r}{P}\right)= \frac{\gamma(a,\frac{r}{Pb})}{\Gamma(a)},\label{eq:op_gam}
\end{align}    
\begin{figure}[t!]
	\begin{center}
		\subfloat[]{\includegraphics[width=45mm,height=45mm]{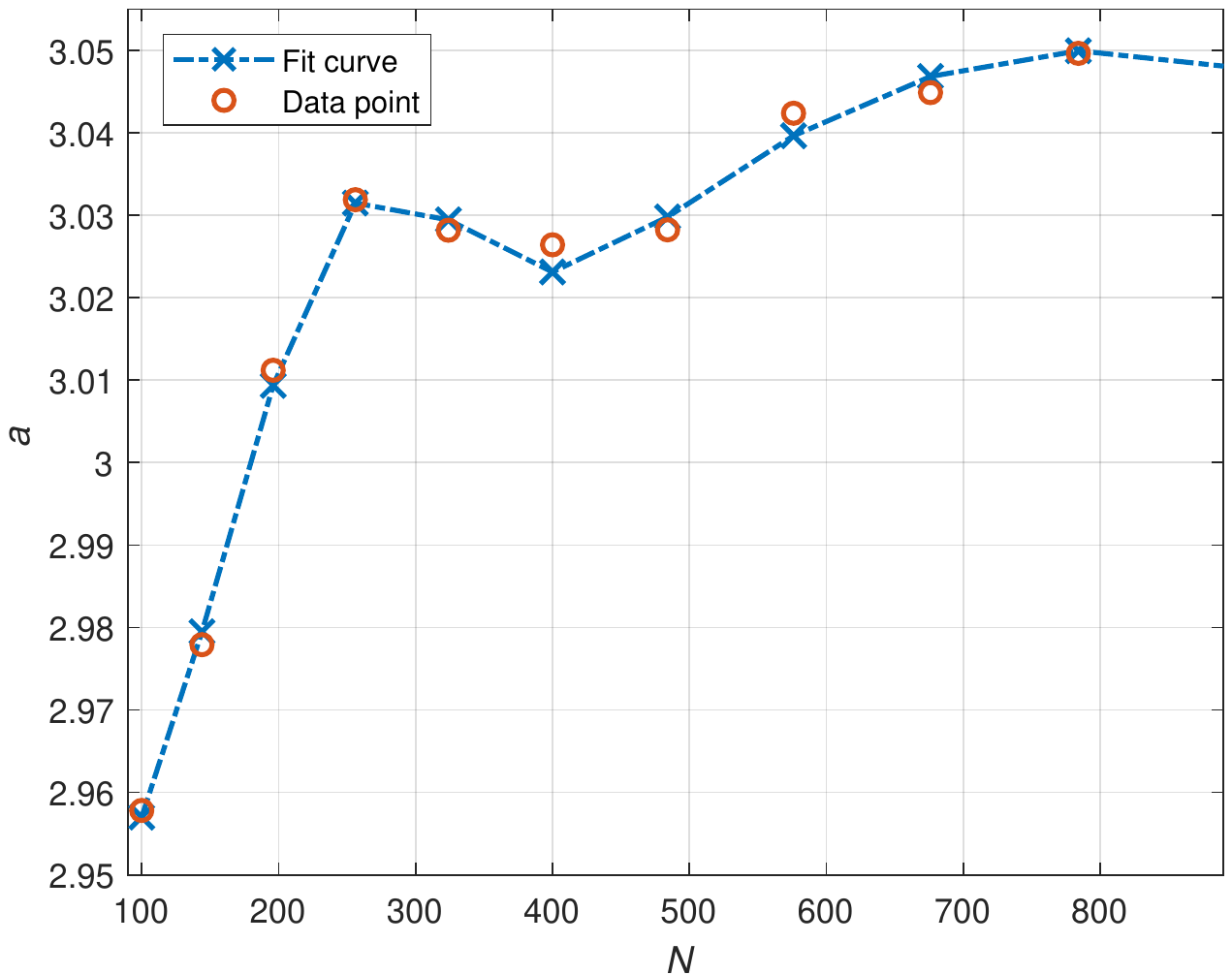}}
		\subfloat[]{\includegraphics[width=45mm,height=45mm]{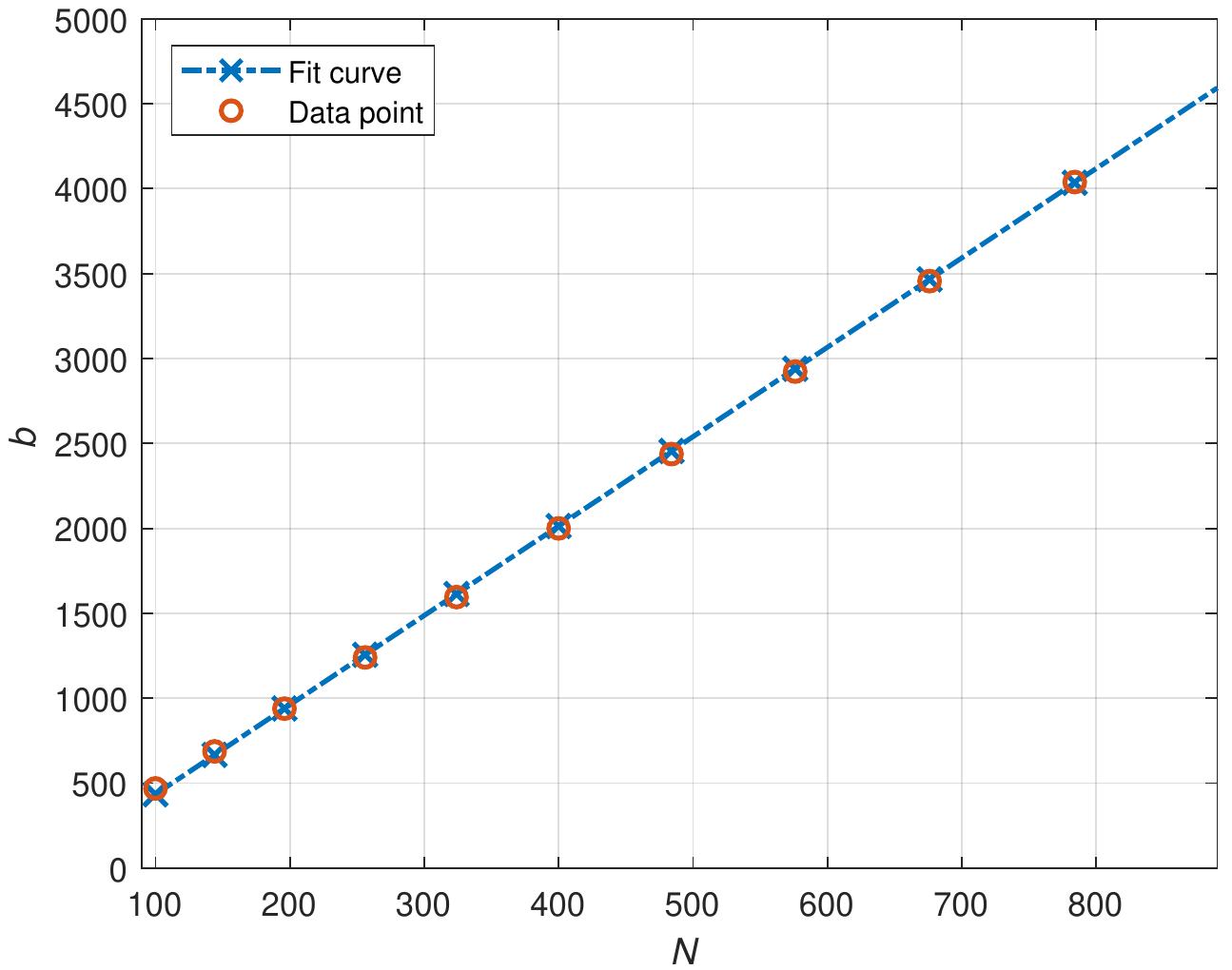}}
		\caption{Fitting Gamma distribution parameters, (a) the shape parameter $a$, and (b) the scale parameter $b$, for $T_2=4T_1$ ($m=4$) and $\rho=0$ dB.}
		\label{fig:a_b_fit}
	\end{center}
	\vspace{-0.5cm}
\end{figure}
\hspace{-0.2cm}where $\Gamma(\cdot)$ and $\gamma(\cdot)$  are the complete and lower incomplete gamma functions, respectively, with  $a$ and $b$ denoting the shape and scale parameters of Gamma distribution, respectively. Furthermore, we used the Curve Fitter Tool in MATLAB to obtain $a$ and $b$ as functions of $N$, as shown in Fig. \ref{fig:a_b_fit}. In particular, we obtained $a=a_1e^{-((N-b_1)/c_1)^2} + a_2e^{-((N-b_2)/c_2)^2}$, with $a_1 = 3.05, b_1 =799.6, c_1 = 3833, a_2 =  0.04247, b_2 =  247.4, c_2 = 109.5$, and $b=p_1N + p_2$, with $p_1=5.262$ and $p_2 = -90.97$.
\begin{table}[t]
	\caption{Shape and scale parameters for different $N$ values.}
	\label{tab:para}
	\begin{center}		
		\begin{tabular}{ |c|c|c|c|} 
			\hline
			$N$ & $a$ & $b$& Standard error for $a$, $b$\\ 
			\hline
			$49$& $2.853$ & $229$ & $0.00382299, 0.336458$\\ 
			\hline
	    	$100$& $2.955$ & $470.391$ & \hspace{-0.1cm}$0.00396615, 0.68805$\\ 
		    \hline
		    $225$& $3.01$ & $1089.9$& \hspace{-0.2cm}$0.0040444,1.59304$\\ 
		    \hline
		\end{tabular}
	\end{center}
	\vspace{-0.5cm}
\end{table}
\section{Simulation Results}\label{sec:emi-sim}
This section presents comprehensive computer simulations to examine the performance of both the proposed and benchmark schemes in terms of average SINR and OP, under different system settings. The considered benchmark is the classical passive beamforming scheme, where the RIS phase shifts are adjusted to remove the overall S-RIS-D channel phases, as given in \eqref{eq:rx2}, at all time slots. In particular, we consider the simulation parameters considered in \cite{EMI_Emil}, where $\sigma_w^2=-114$ dBm, $\beta_1=-48$ dB, $\beta_2=-38$ dB. Furthermore, as in \cite{EMI_Emil}, we use $\rho$ to control the signal to the EMI power ratio, where $\rho\in\{0,5,10\}$ dB. The RIS spatial correlation model in \cite{spatial-corr} is adopted with $\lambda/2$ separation between elements, where $\lambda$ is the wavelength associated with the operating frequency $1.8$ GHz. The considered CI of $\mathbf{h}_2$ is $T_2=4T_1$ ($m=4$) unless otherwise stated. Finally, for all of our simulations, we provide the SINR averaged over $m$ time slots ($\text{SINR}_m$); that is, the SINR performance over a full CI of $\mathbf{h}_2$.
\begin{figure}[t]
	\centering
	\includegraphics[width=60mm]{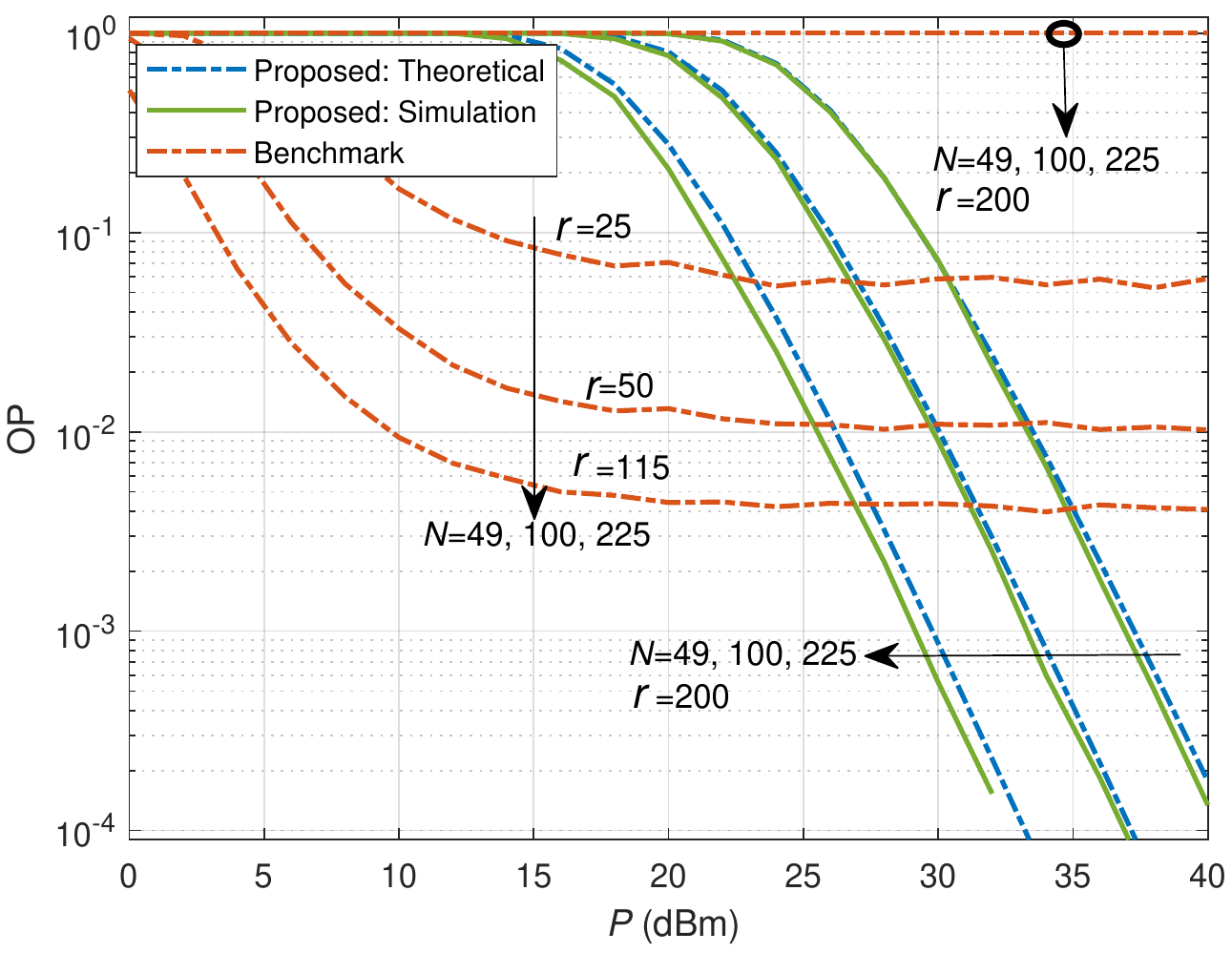}
	\caption{Outage probability performance for $\rho=0$ dB and different $N$ values.}
	\label{fig:OP}
\end{figure}

For the spatial correlation matrix $\mathbf{R}$, its entries are given as $[\mathbf{R}]_{k,\tilde{k}}=\text{sinc}\left(2\norm{\mathbf{u}_k-\mathbf{u}_{\tilde{k}}}/\lambda\right),\; k,\tilde{k}=1,...,N$ \cite{spatial-corr},
where $\text{sinc}(a) = \text{sin}(\pi a)/(\pi a)$ is the sinc function, $\mathbf{u}_k=[0,\;i(k)d_V,\tilde{i}(k)d_H\;]^T$, with $i(k)$, $\tilde{i}(k)$, $d_V$, and $d_H$ are the horizontal index, vertical index, length, and width of the element $k$, respectively, 
$i(k)=\text{mod}(k-1,N_H)$, $\tilde{i}(k)=\left\lfloor(k-1)/N_H\right\rfloor$.

In Fig. \ref{fig:OP}, we provide the outage probability performance for different RIS sizes, where the theoretical and simulation curves are shown to have a close match to each other, using the fitting parameters given in Table \ref{tab:para}. It can be seen that for an $\text{SINR}_m$ threshold $r=200$, the benchmark scheme has an OP of unity for all considered $N$ values, while the proposed scheme has a superior performance that improves with increasing $P$ and $N$. Furthermore, to clarify the performance of the benchmark scheme further, we consider different threshold ($r$) values, since considering a single threshold value does not accurately reflect the performance for different $N$ values due to the sensitivity of the OP to varying $N$. Accordingly, although the different threshold values of the benchmark scheme are much lower than the one of the proposed scheme, yet, it can be seen that at high $P$ region, the proposed scheme has a superior OP performance. This is because, unlike the proposed scheme that eliminates the EMI, the performance of the benchmark scheme gets saturated due to the EMI, where asymptotically, the average SINR approaches to a constant value as $P$ increases.

\begin{figure}[t]
	\centering
	\includegraphics[width=60mm]{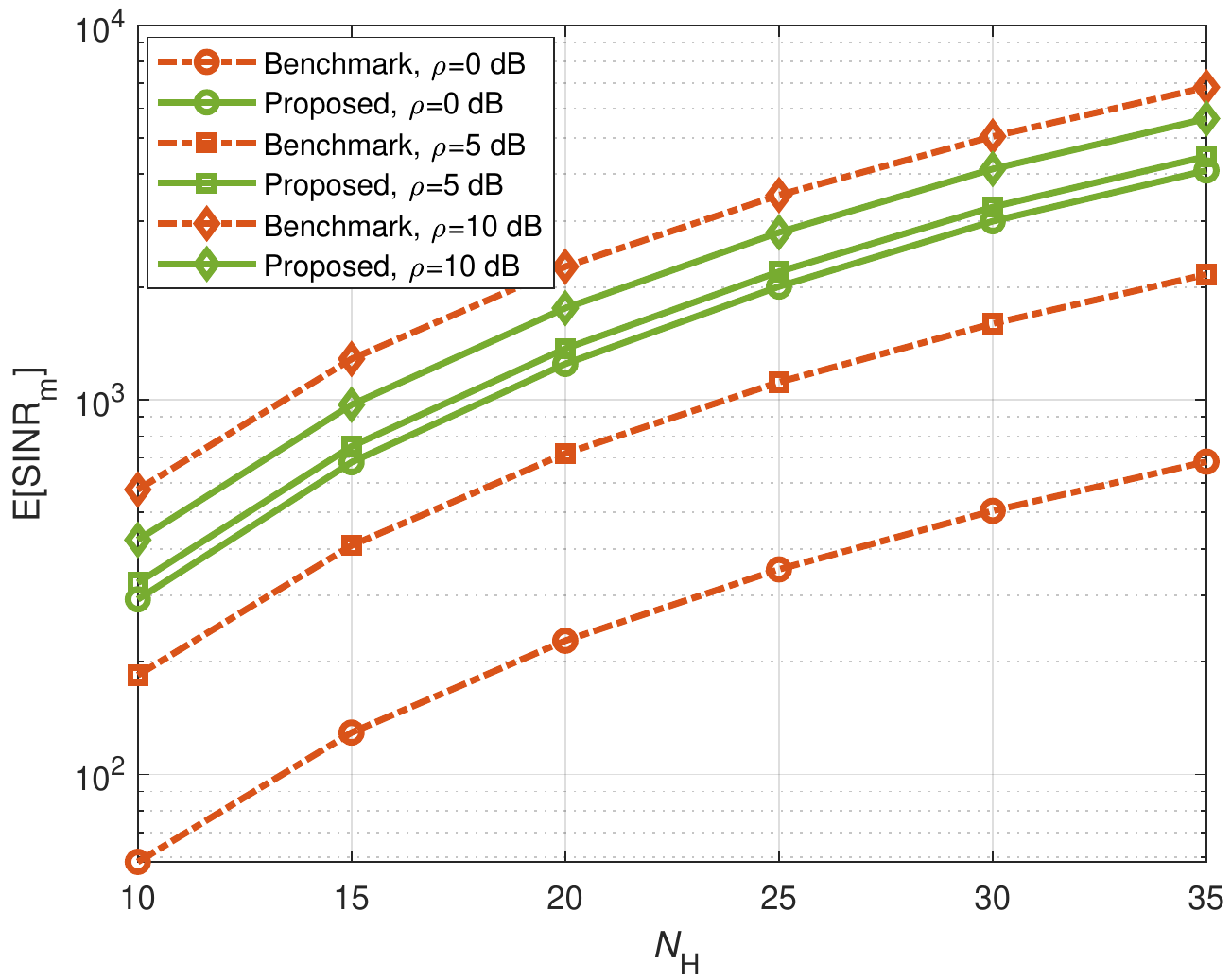}
	\caption{Average SINR performance for different RIS sizes and EMI power levels, with $P=23$ dBm \cite{EMI_Emil}.}
	\label{fig:N_effect}
	\vspace{-0.5cm}
\end{figure}

In Fig. \ref{fig:N_effect}, we provide the average SINR versus different RIS sizes and EMI power levels, where the x-axis corresponds to the number of elements per RIS dimension, $N_H=\sqrt{N}$. It can be seen that increasing the RIS size (x-axis) does not affect the SINR gap between the proposed and benchmark scheme, as increasing the RIS size will both increase the beamforming gain and the amount of the EMI reflected to D. Furthermore, it can be noted that our proposed EMI cancellation scheme is more effective when the EMI power is higher, where removing the EMI has a better impact on the SINR compared to boosting the received signal power through the passive beamforming. On the other side, the benchmark scheme performs better under low levels of EMI power, where the EMI has less impact with the dominating AWGN power and thus, boosting the signal power provides a better impact on the SINR than removing the EMI.
\begin{figure}[t]
	\centering
	\includegraphics[width=60mm]{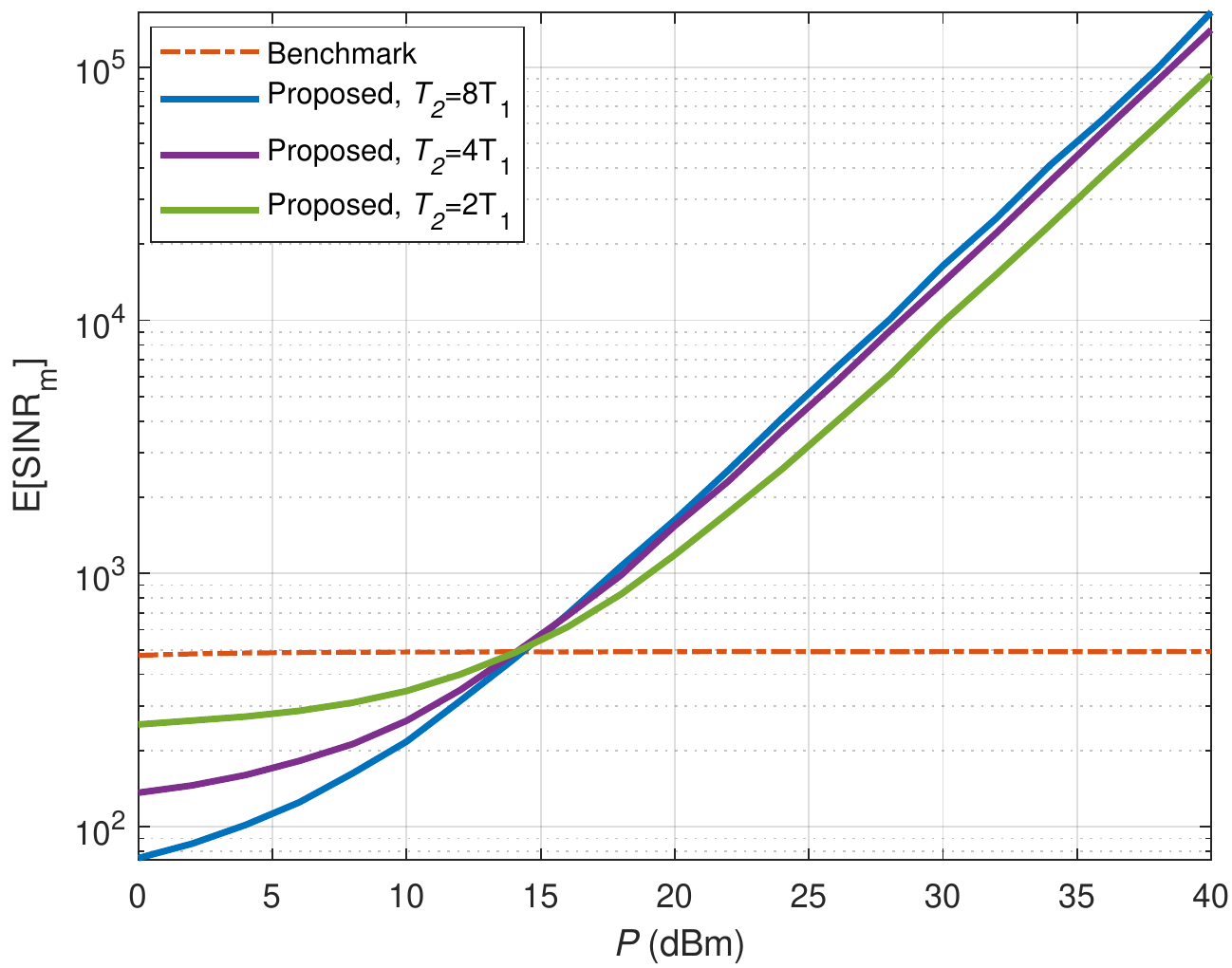}
	\caption{Average SINR performance under increasing CI length ($T_2$) with $N=900$ and $\rho=0$ dB.}
	\label{fig:T_effect}
\end{figure}
\begin{figure}[h]
	\centering
	\includegraphics[width=60mm]{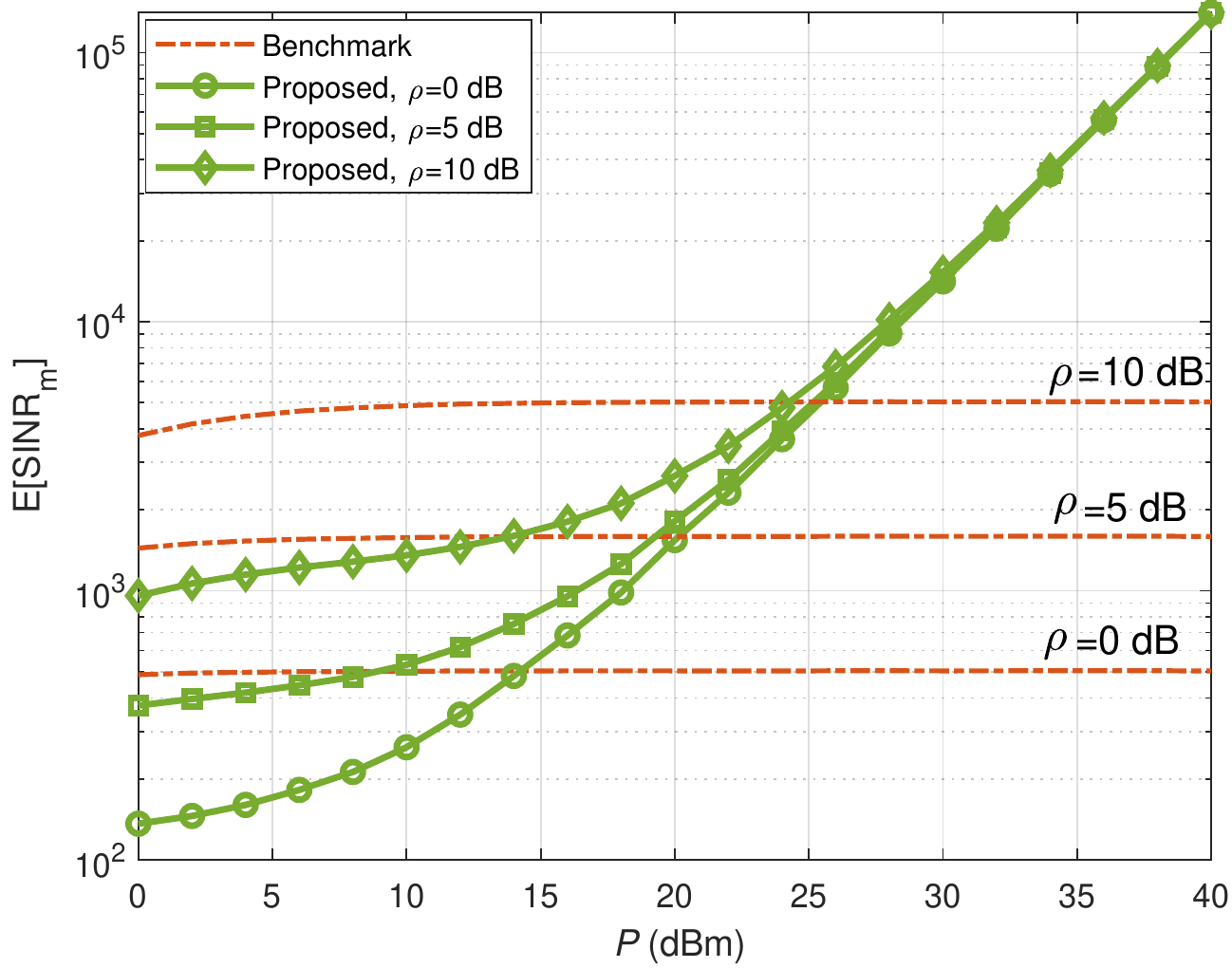}
	\caption{Average SINR performance against EMI power level with $N=900$.}
	\label{fig:Rho_effect}
		\vspace{-0.5cm}
\end{figure}

In Fig. \ref{fig:T_effect}, we show the system performance as the CI of $\mathbf{h}_2$ changes, which means that the CI of $\mathbf{n}$ changes too as the CI of the latter is greater than $T_2$, as shown in Fig. \ref{fig:CIs}. It can be seen that, as $T_2/T_1$ ratio increases, a better SINR performance is achieved by the proposed scheme, where the SINR values at $t^{'}$ dominate the one obtained at the first time slot; thus, the overall average over $m$ increases. Furthermore, the crossing point $P=14$ dBm corresponds to the critical point at which the EMI power starts dominating the AWGN power, where for $\rho=0$ dB we have $\sigma^2=\beta_1 P$ at the RIS side, which eventually becomes $\sigma^2=\beta_1 \beta_2 P=-100 \;\text{dBm}>\sigma_w^2$ at D side. Note that this gives an advantage to our proposed scheme, which removes the EMI and outperforms the benchmark scheme. On the other side, before this critical point, the AWGN power dominates, and the benchmark scheme performs better as it boosts the SNR.

Finally, in Fig. \ref{fig:Rho_effect}, the effect of the EMI power is investigated for $N=900$, where a large RIS size is necessary here to reveal the impact of the EMI \cite{EMI_Emil}. It can be seen that the proposed scheme performs better as $\rho$ decreases (EMI power increases). Also, the performance of the benchmark scheme gets saturated in a fast manner while the proposed scheme keeps improving with $P$. Overall, it can be noted that, unless the main signal power is much stronger than the EMI, it is better to adjust the RIS phase shifts to eliminate the EMI instead of steering (beamforming) the main signal to D.
\section{Conclusion}\label{sec:emi-conc}
In this paper, we have proposed a novel EMI cancellation scheme by trading off the RIS passive beamforming in favor of eliminating the EMI by properly designing the RIS phase shifts over different time slots. Compared to the benchmark scheme that ignores the EMI effects, the proposed scheme is shown to achieve better performance in terms of average SINR and outage probability, particularly when the EMI power is comparable to the main signal power reflected from the RIS surface. For future research, extending the proposed scheme to a MIMO setup seems a promising direction.
%
\bibliographystyle{IEEEtran}
\bibliography{IEEEabrv,Bibliography}

\begin{thebibliography}{10}
\providecommand{\url}[1]{#1}
\csname url@samestyle\endcsname
\providecommand{\newblock}{\relax}
\providecommand{\bibinfo}[2]{#2}
\providecommand{\BIBentrySTDinterwordspacing}{\spaceskip=0pt\relax}
\providecommand{\BIBentryALTinterwordstretchfactor}{4}
\providecommand{\BIBentryALTinterwordspacing}{\spaceskip=\fontdimen2\font plus
\BIBentryALTinterwordstretchfactor\fontdimen3\font minus
  \fontdimen4\font\relax}
\providecommand{\BIBforeignlanguage}[2]{{%
\expandafter\ifx\csname l@#1\endcsname\relax
\typeout{** WARNING: IEEEtran.bst: No hyphenation pattern has been}%
\typeout{** loaded for the language `#1'. Using the pattern for}%
\typeout{** the default language instead.}%
\else
\language=\csname l@#1\endcsname
\fi
#2}}
\providecommand{\BIBdecl}{\relax}
\BIBdecl

\bibitem{Transmission_conference}
E.~Basar, ``Transmission through reconfigurable intelligent surfaces: A new
  frontier in wireless communications,'' in \emph{Proc. European Conf. Netw.
  Commun. (EuCNC)}, Valencia, Spain, June 2019.

\bibitem{AK1}
A.~Khaleel and E.~Basar, ``Reconfigurable intelligent surface-empowered {MIMO}
  systems,'' \emph{IEEE Syst. J.}, vol.~15, no.~3, pp. 4358--4366, Aug. 2021.

\bibitem{AK2}
------, ``A novel {NOMA} solution with {RIS} partitioning,'' \emph{IEEE J. Sel.
  Topics Signal Process.}, vol.~16, no.~1, pp. 70--81, Jan. 2022.

\bibitem{mahmoud}
M.~Aldababsa, A.~Khaleel, and E.~Basar, ``Simultaneous transmitting and
  reflectingintelligent surfaces-empowered {NOMA} networks,'' Oct. 2021.
  [Online]. Available: arXiv:2110.05311.

\bibitem{EMI_Emil}
A.~de~Jesus~Torres, L.~Sanguinetti, and E.~Björnson, ``Electromagnetic
  interference in {RIS}-aided communications,'' \emph{IEEE Wireless Commun.
  Lett.}, vol.~11, no.~4, pp. 668--672, Apr. 2022.

\bibitem{ris-snr-scale}
Q.~Wu and R.~Zhang, ``Towards smart and reconfigurable environment: Intelligent
  reflecting surface aided wireless network,'' \emph{IEEE Commun. Mag.},
  vol.~58, no.~1, pp. 106--112, Jan. 2020.

\bibitem{snr-sqr-law}
E.~Basar and H.~V. Poor, ``Present and future of reconfigurable intelligent
  surface-empowered communications [perspectives],'' \emph{IEEE Signal Process.
  Mag.}, vol.~38, no.~6, pp. 146--152, Nov. 2021.

\bibitem{emi-relay}
A.~d.~J. Torres, L.~Sanguinetti, and E.~Björnson, ``Intelligent reconfigurable
  surfaces vs. decode-and-forward: What is the impact of electromagnetic
  interference?'' in \emph{2022 IEEE 23rd Int. Workshop Signal Process.
  Advances Wireless Commun. (SPAWC)}, Jul. 2022, pp. 1--5.

\bibitem{emi-urllc}
G.~S. Chandra, R.~K. Singh, S.~Dhok, P.~K. Sharma, and P.~Kumar, ``Downlink
  {URLLC} system over spatially correlated {RIS} channels and electromagnetic
  interference,'' \emph{IEEE Wireless Commun. Lett.}, vol.~11, no.~9, pp.
  1950--1954, Sep. 2022.

\bibitem{emi-phys}
J.~David Vega-Sánchez, G.~Kaddoum, and F.~J. López-Martínez, ``Physical
  layer security of {RIS}-assisted communications under electromagnetic
  interference,'' \emph{IEEE Commun. Lett.}, vol.~26, no.~12, pp. 2870--2874,
  Dec. 2022.

\bibitem{emi-multipair}
S.~R. Kudumala, A.~K. Dubey, P.~Gupta, S.~Gupta, and E.~Sharma, ``Hardware
  impaired {RIS} assisted multipair {FD} communication with spatial
  correlation,'' \emph{IEEE Commun. Lett.}, vol.~26, no.~9, pp. 2200--2204,
  Sep. 2022.

\bibitem{RIS-Interf}
T.~Jiang and W.~Yu, ``Interference nulling using reconfigurable intelligent
  surface,'' \emph{IEEE J. Sel. Areas Commun.}, vol.~40, no.~5, pp. 1392--1406,
  May 2022.

\bibitem{spatial-corr}
E.~Björnson and L.~Sanguinetti, ``Rayleigh fading modeling and channel
  hardening for reconfigurable intelligent surfaces,'' \emph{IEEE Wireless
  Commun. Lett.}, vol.~10, no.~4, pp. 830--834, Apr. 2021.

\bibitem{ris-position}
E.~Ibrahim, R.~Nilsson, and J.~van~de Beek, ``On the position of intelligent
  reflecting surfaces,'' in \emph{2021 Joint European Conf. Netw. Commun. \& 6G
  Summit (EuCNC/6G Summit)}, 2021, pp. 66--71.

\bibitem{ris-position2}
E.~Basar, I.~Yildirim, and F.~Kilinc, ``Indoor and outdoor physical channel
  modeling and efficient positioning for reconfigurable intelligent surfaces in
  mmwave bands,'' \emph{IEEE Trans. Commun.}, vol.~69, no.~12, pp. 8600--8611,
  Dec. 2021.

\bibitem{H2_slow1}
H.~Liu, X.~Yuan, and Y.-J.~A. Zhang, ``Matrix-calibration-based cascaded
  channel estimation for reconfigurable intelligent surface assisted multiuser
  {MIMO},'' \emph{IEEE J. Sel. Areas Commun.}, vol.~38, no.~11, pp. 2621--2636,
  Nov. 2020.

\bibitem{H2_slow2}
C.~Hu, L.~Dai, S.~Han, and X.~Wang, ``Two-timescale channel estimation for
  reconfigurable intelligent surface aided wireless communications,''
  \emph{IEEE Trans. Commun.}, vol.~69, no.~11, pp. 7736--7747, Nov. 2021.

\bibitem{EMI_slow}
R.~Long, Y.-C. Liang, H.~Guo, G.~Yang, and R.~Zhang, ``Symbiotic radio: A new
  communication paradigm for passive internet of things,'' \emph{IEEE Internet
  Things J.}, vol.~7, no.~2, pp. 1350--1363, Feb. 2020.

\end{thebibliography}
\end{document}